\documentclass[twocolumn,aps,prl,showpacs]{revtex4-1}
\usepackage{epsfig}
\usepackage{graphics}
\usepackage{amssymb}
\usepackage{amsmath}

\newcommand{\hc}{\hat{c}}

\newcommand{\ha}{\hat{a}}

\newcommand{\hm}{\hat{m}}

\newcommand{\hb}{\hat{b}}

\begin{document}

\title{Erasing Quantum Distinguishability via Single-Mode Filtering}

\author{Monika Patel,$^1$ Joseph B. Altepeter,$^2$ Yu-Ping Huang,$^2$ Neal N. Oza,$^2$ and Prem Kumar$^{1,2}$}
\affiliation{Center for Photonic Communication and Computing, \\
$^1$Department of Physics and Astronomy, $^2$Department of Electrical Engineering and Computer Science,\\
Northwestern University, 2145 Sheridan Road, Evanston, IL, 60208-3118, USA}

\begin{abstract}
Erasing quantum-mechanical distinguishability is of fundamental interest and also of practical importance, particularly in subject areas related to quantum information processing. We demonstrate a method applicable to optical systems in which single-mode filtering is used with only linear optical instruments to achieve quantum indistinguishability. Through ``heralded'' Hong-Ou-Mandel interference experiments we measure and quantify the improvement of indistinguishability between single photons generated via spontaneous four-wave mixing in optical fibers. The experimental results are in excellent agreement with predictions of a quantum-multimode theory we develop for such systems, without the need for any fitting parameter.
\end{abstract}
\maketitle

Quantum indistinguishability is inextricably linked to several fundamental phenomena in quantum mechanics, including interference, entanglement, and decoherence \cite{ent, decoh1, decoh2}. For example, only when two photons are indistinguishable can they show strong second-order interference \cite{indis}. From an applied perspective, it forms the basis of quantum key distribution \cite{bb84}, quantum computing \cite{KniLafMil01}, quantum metrology \cite{quantummetrology}, and many other important applications in modern quantum optics. In practice, however, the generation and manipulation of quantum-mechanically indistinguishable photons is quite challenging, primarily due to their coupling to external degrees of freedom.

In this Letter, we experimentally investigate a pathway to erasing quantum distinguishability by making use of the Heisenberg uncertainty principle. This method, although designed specifically for optical systems, might be generalizable to other physical systems, including those of atoms and ions. It uses a filtering device that consists of only linear optical instruments, which in our present rendering is a temporal gate followed by a spectral filter. The gate's duration $T$ and the filter's bandwidth $B$ (in angular-Hertz) are chosen to satisfy $BT<1$ so that any photon passing through the device loses its temporal (spectral) identity as required by the Heisenberg uncertainty principle. In this sense, the device behaves as a single-mode filter (SMF) that passes only a single electromagnetic mode of certain temporal profile while rejecting all other modes. Hence, applying such a SMF to distinguishable single photons can produce output photons that are indistinguishable from each other \cite{HuaAltKum10, HuaAltKum11}. Our calculations show that for appropriate parameters very high levels of quantum indistinguishability can be achieved with use of the SMF, while paying a relatively low cost in terms of photon loss. This method is superior to using tight spectral or temporal filtering alone for similar purposes \cite{ZeiHorWei97, FioVosSha02}, where the photon loss is much higher. In fact, in Refs.~\cite{HuaAltKum10, HuaAltKum11} we have shown that the use of a SMF can significantly improve the performance of heralding-type single-photon sources made from optical fibers or crystalline waveguides \cite{Heralded-Single-Photon-SPDC86,Single-Photon-PCF05, Single-Photon-Fiber09, Single-Photon-PDC-99}.

In our experiment, pairs of signal and idler photons are generated in two separate optical-fiber spools via spontaneous four-wave mixing. By detecting the idler photons created in each spool, we herald the generation of their partner (signal) photons. To quantify their indistinguishability, we mix the signal photons generated separately from the two spools on a 50:50 beamsplitter and perform Hong-Ou-Mandel (HOM) interference measurements. We find that the HOM visibility is quite low when the signal photons have a temporal length $T>1/B$, owing to the presence of photons with many distinguishable degrees of freedom. However, when $T<1/B$, for which a SMF is effectively realized, a much higher HOM visibility is obtained. This result clearly shows that the SMF can be used to erase the quantum distinguishability of single photons. To quantitatively examine the degree of improvement, we develop a comprehensive theoretical model of light scattering and detection in optical fiber systems, taking into account multi-pair emission, Raman scattering, transmission loss, dark counts, and other practical parameters. The experimental data are in good agreement with predictions of the model without the need for any fitting parameter.

To understand our approach for erasing quantum distinguishability, we consider amplitude profiles $f(t)$ and $h(\omega)$ for the time gate and the spectral filter, respectively. The number operator for output photons is then given by
$\hat{n}=\frac{1}{(2\pi)^2}\int d\omega d\omega' \kappa(\omega,\omega')\ha^\dag(\omega)\ha(\omega')$ \cite{PrSp61,ZhuCav90}, where $\ha(\omega)$ is the annihilation operator for the
incident photons of angular-frequency $\omega$, satisfying
$[\ha(\omega),\ha^\dag(\omega')]=2\pi\delta(\omega-\omega')$.
$\kappa(\omega,\omega')=\int dt~ h^\ast(\omega) h(\omega') |f(t)|^2
e^{i(\omega-\omega')t}$ is a Hermitian spectral correlation
function, which can be decomposed onto a set of
Schmidt modes as
$
    \kappa(\omega,\omega')=\sum^{\infty}_{j=0} \chi_{j} \phi^\ast_{j}(\omega)\phi_{j}(\omega'),
$
where $\{\phi_{j}(\omega)\}$ are the
mode functions satisfying $\int d\omega
\phi^\ast_{j}(\omega)\phi_{k}(\omega)=2\pi\delta_{j,k}$ and
$\{\chi_{j}\}$ are the decomposition coefficients satisfying
$1\ge \chi_0 >\chi_1>...\ge 0$. Introducing an infinite set
of mode operators via $\hc_{j}=\frac{1}{2\pi}\int d\omega
\ha(\omega) \phi_{ j}(\omega)$ ($j=0,1,...$) that satisfy
$[\hc_{j},\hc_{k}^\dag]=\delta_{jk}$, the output operator for the filtering device can be
rewritten as
\begin{equation}
    \hat{n}=\sum^{\infty}_{j=0} \chi_j~ \hc^\dag_{j} \hc_{j}.
\end{equation}
This result indicates that $\{\phi_{j}(\omega)\}$ have an intuitive physical
interpretation: as ``eigenmodes'' with eigenvalues $\{\chi_{j}\}$ of the filtering device. In this physical
model, the filtering device projects incident photons onto
the eigenmodes, each of which are passed with a probability
given by the eigenvalues. Specifically, for $\chi_{0}\sim 1$ and
$\chi_{j\neq 0} \ll 1$ (achievable with an appropriate
choice of spectral and temporal filters, as shown below)
only the fundamental mode is transmitted while all the other
modes are rejected. In this way, truly \emph{single-mode} filtering
can be achieved.  Combined with single-photon detectors, this can be
extended to a single-mode, single-photon detection system.
Regardless of the type of spectral and temporal filters used to
achieve this kind of single-mode filtering, such a system is
capable of separating photons which, even though they may exist in the
same spectral band and the same time-bin, have different mode structures.

\begin{figure}
\centering \epsfig{figure=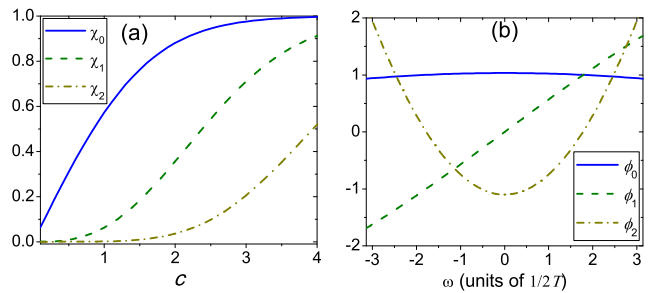, width=8.5cm} \caption{(Color online) (a) $\chi_0$, $\chi_1$, and $\chi_2$ as functions of $c$. (b) Plots of $\phi_0$, $\phi_1$, and $\phi_2$ for $B=\pi/T$ (corresponding to $c=\pi/4$).  \label{fig1}}
\end{figure}

As an example, in Fig.~\ref{fig1}(a) we plot $\chi_{0},\chi_{1},\chi_{2}$
versus $c\equiv BT/4$ for a rectangular-shaped spectral filter with bandwidth $B$ and a rectangular-shaped time window of duration $T$ \cite{PrSp61,SasSuz06}. For $c<1$, we have $\chi_0\approx 1$ whereas
$\chi_1,\chi_2\ll 1$, giving rise to approximately single-mode
filtering.  Note that this behavior is true for any $B$, as
long as $T<4/B$. In other words, $\{\chi_j\}$ depend only on the
product of $B$ and $T$, rather than on their specific values.
Consequently, even a broadband filter can lead to a single-mode measurement over
a sufficiently short detection window, and vice-versa. To understand
this, consider the case where a detection event announces the
arrival of a signal photon at an unknown time within the window
$T$. In the Fourier domain, this corresponds to a detection
resolution of $1/T$ in frequency. Given $c<1$ or $1/T>B/4$,
the detector is thus unable to, even in principle, reveal the
frequency of the signal photon. Therefore, the signal photon
is projected onto a quantum state in a coherent superposition of
frequencies within $B$ \cite{HuaAltKum11}. This can be seen
in Fig.~\ref{fig1}(b), where the fundamental detection mode has
a nearly flat profile over the filter band $[-B/2, B/2]$. Lastly,
since $T<4/B$ is required, the pass probability of the fundamental
mode will be sub-unity, but not significantly less than one.

To verify this theory of erasing quantum distinguishability via single-mode filtering, we perform a heralded two-photon interference experiment \cite{gisin03, zeilingerentswp09, rarity07, takesue07} in both multimode ($c>1$) and single-mode ($c<1$) regimes. Hong-Ou-Mandel interference between two photons originating from independent
photon-pair sources provides a test of indistinguishability. Appropriate choices of wavelength-division multiplexers (spectral filters which select $B$) and the width of pulses pumping the photon-pair sources (which effectively sets the temporal window $T$ in which photon pairs are born) allow a transition from the single-mode to the multimode regime.
\begin{figure}
\centering
\epsfig{figure=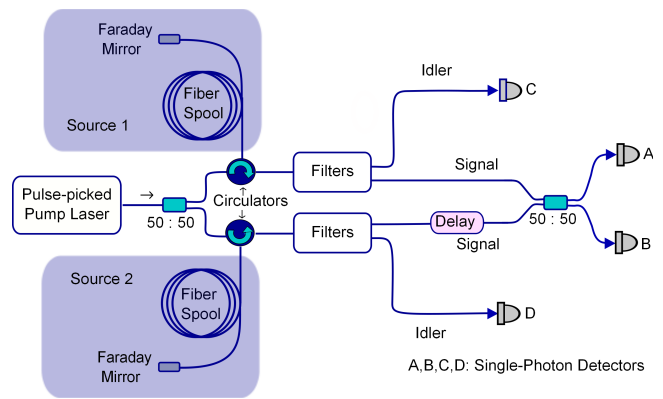, width=8.5cm}
\caption{(Color online) Experimental setup. A,B,C,D: single-photon detectors.}
\label{experimentalsetup}
\end{figure}

The experimental setup is shown in Fig.~\ref{experimentalsetup}. Both heralded photon-pair sources are pumped using the same system, consisting of 50-MHz repetition-rate pulses carved from the output of either a continuous-wave (CW) laser (for the multimode heralding experiment) or a mode-locked laser (for the single-mode heralding experiment). The pulse-carver is an optical amplitude modulator (EOSPACE, Model AK-OK5-10) driven by the output of a 20-Gbps 2:1 selector (Inphi, Model 20709SE), which is clocked at 50 MHz by an electrical signal source that also triggers the single-photon detectors (NuCrypt, Model CPDS-4) used in the experiment. The carved pulses are then amplified and fed to a 50:50 fiber splitter. Each output branch of the splitter leads to a four-wave-mixing (FWM) fiber spool (500 m of standard single-mode fiber cooled to 77 K) in a Faraday-mirror configuration \cite{HuaAltKum09}. The Faraday mirror effectively doubles the length of fiber available for four-wave mixing while simultaneously compensating for any polarization changes which may occur in the spooled fiber. The signal and idler photons are created via spontaneous four-wave mixing. Along with the residual pump photons, they enter two cascaded filtering stages which provide $\approx$100-dB of isolation. The filtered signal and idler photons then pass through fiber polarization controllers (not shown in Fig.~\ref{experimentalsetup}) and the signal photons are led to the two input ports of an in-fiber 50:50 coupler. Adjusting the polarization controllers and careful temporal alignment with use of a variable delay stage in the path of one of the signal photons ensures that the signal photons arriving at the 50:50 coupler are identical in all degrees of freedom: polarization, spectral/temporal, and spatial. Note that even though these signal photons are identical, they may still be partially or completely distinguishable (particularly in the multimode regime described above). This distinguishability may arise from entanglement with \emph{different} idler photons (heralds) or from the presence of background photons that originate in the FWM fiber owing to Raman scattering. Four InGaAs-based single-photon detectors are used to count photons, one each at the outputs of the idler arms and the 50:50 coupler. These detectors are gated at 50-MHz repetition rate synchronous with the arrival of photons and have a dark-count probability of $1.6\times10^{-4}$ per pulse. Their quantum efficiencies are approximately 20\%. The delay stage is used to vary the temporal overlap of the signal photons while the photon counts are recorded.

In the multimode experimental configuration, where a CW laser (Santec, model TSL-210V) is used as the pump, the temporal duration of the carved pulses is specified by the width of the electrical pulses provided to the modulator, which is measured to be 100~ps, giving $T=10^{-10}$~s. The signal and idler filters each consist of a free-space diffraction-grating filter [full-width at half-maximum (FWHM) $\approx$ 0.14 nm] followed by a dense wavelength-division-multiplexing (DWDM) filter (FWHM $\approx$ 0.4 nm). The resulting optical transmission spectra are shown in Fig.~3(a), from which the effective bandwidth of the signal and idler filters is determined to be approximately 0.14 nm. In units of frequency, this gives $B/2\pi=24.6$ GHz so that $BT=2.5\times2\pi$. Therefore, $c=3.8$ and from Fig.~\ref{fig1}, $\chi_0\approx1$, $\chi_1\approx0.9$ and $\chi_2\approx 0.5$. Because $\chi_1$ and $\chi_2$ are not neglible, this case corresponds to a multimode measurement.

\begin{figure}
\centering
\epsfig{figure=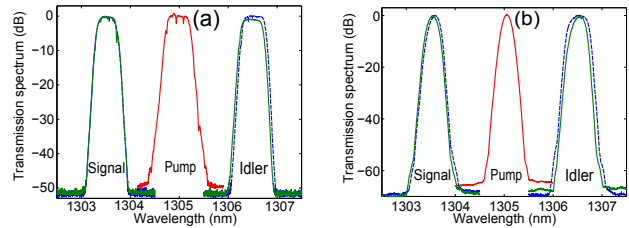, width=8.7cm}
\caption{(Color online) Optical transmission spectra for each stage of the pump, signal, and idler filters. Each filter is composed of two separate stages and  provides $>$100 dB of isolation. The stages are either fiber-coupled free-space double-pass transmission-grating filters or DWDM filters (custom-made by AC Photonics, Inc.). (a) Spectra for the pump filter (formed by two grating filters) and the signal and idler filters (each formed by a grating and a DWDM filter) used in the multimode experiment. (b) Spectra for the pump, signal, and idler filters used in the single-mode experiment, each formed by two DWDM filters centered at the respective wavelength. In both plots, solid (green) and dashed (blue) traces correspond to the filters used in source 1 and source 2, respectively.
\label{filters}}
\end{figure}

In the single-mode experimental configuration, a 10-GHz mode-locked laser (U2T, model TMLL1310) emitting a train of 2-ps duration, transform-limited pulses is used as the pump. The signal and idler photons along with the pump pulses are each filtered by two stages of DWDM filters. The resulting optical transmission spectra of these filters are shown in Fig.~3(b). The bandwidth of the pump filter is measured to be $68.3$ GHz, from which the pump-pulse width and thus the effective $T$ is derived to be $6.4$ ps. The bandwidths of the signal and the idler filters, on the other hand, are both approximately 0.4 nm, which give $BT\approx 0.4\times 2\pi$ or $c=0.7$. In this case, $\chi_0\approx0.4$, and $\chi_1$ and $\chi_2$ are nearly zero, giving rise to a single-mode measurement.

\begin{figure}
\centering
\epsfig{figure=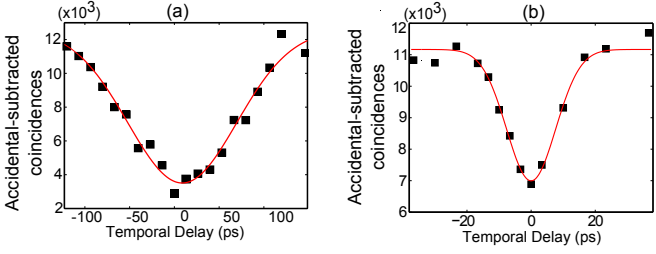, width=9cm}
\caption{(Color online) (a) Accidental-subtracted A-B coincidences recorded per $5\times 10^7$ pump pulses as a function of the relative delay between signal photons in the multimode configuration; and (b) in the single-mode configuration. Error bars are of the same size as the data markers. The red curve is a Gaussian least-square fit to the data.}
\label{trues}
\end{figure}

In practice it is experimentally convenient to analyze the behavior of non-heralded two-photon coincidence counts to precisely path-match the two signal arms. This is because there are many more twofold coincidences than fourfold coincidences in the system, which allows us to study the quantum interference effect with much smaller error bars and a much shorter measurement time. To this end we define a twofold coincidence count to be when detectors A and B (c.f. Fig. 1) fire in the same time slot. We define a twofold accidental-coincidence count to be when detectors A and B fire in adjacent time slots. Finally, we define a fourfold coincidence, the quantity of primary experimental interest, to occur when all four detectors fire simultaneously in the same time slot. Figure~\ref{trues} shows the variation in accidental-subtracted coincidences on detectors A and B as the relative delay between the signal photons from the two FWM sources is varied.

\begin{figure}
\centering
\epsfig{figure=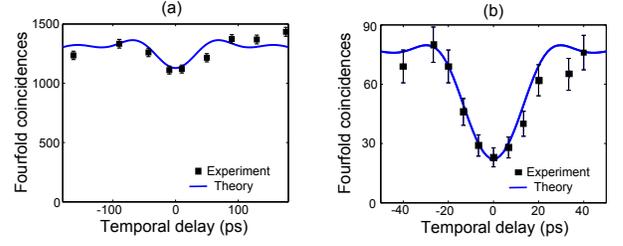, width=8.5cm}
\caption{(Color online) (a) Fourfold coincidence counts per 20 billion pulses recorded as a function of timing delay,
using the continuous-wave laser as pump, and (b) fourfold coincidence counts per 10 billion pulses recorded as a function of timing delay, using the mode-locked laser as pump. In both figures, the error bars are computed following the standard manner of estimating statistical randomness assuming Poissonian distributions of the recorded coincidence and singles counts.}
\label{expresults}
\end{figure}

The recorded fourfold coincidence counts as a function of the relative delay in the heralded HOM interference experiment are plotted in Fig.~\ref{expresults}. For the multimode experimental configuration, as shown in Fig.~\ref{expresults}(a), the interference visibility is only $19\pm2\%$. In contrast, for the single-mode configuration, a high visibility of $72\pm7\%$ is obtained, as shown in Fig.~\ref{expresults}(b). This is the highest HOM interference visibility reported thus far for fiber-based single-photon sources in the telecommunications band. For these results, the transmission efficiencies of the signal and idler photons from their generation site in the FWM spools to the detectors are measured for each arm and are found to be 3.4\% (5.5\%) for the signal arms and 5.0\% (7.0\%) for the idler arms in the multimode (single-mode) configuration. The photon-pair production probabilities per pump pulse are measured to be 12.5\% and 3.9\% for the multimode and single-mode configurations, respectively.

Although these experiments show a clear difference between the single-mode and multimode regimes, the theory of single-mode detection presented above---in the absence of any systematic sources of noise---seems to predict much higher visibilities, particularly for the single-mode experiment where it seems that any entanglement with the idler photons should be eliminated by the SMF. In fact, systematic sources of noise---from multi-pair production, stimulated Raman emission, loss, and dark counts---do significantly affect the results. In order to determine the extent to which these experimental results verify the theory of SMF presented above, it is necessary to create a complete theoretical model of multi-pair production, Raman emission, loss, dark-count noise, and the interference between two real experimental systems. For this goal, we adopt the standard quantum-mechanical description (assuming phase matching and undepleted pump) of light scattering in optical fibers at a few-photon level \cite{FWM-Raman07}:
$  \ha^{r(\ell)}_{s,a}(\omega)= \int d\omega' \alpha(\omega-\omega') \hb^{r(\ell)}_{s,a}(\omega') +i\gamma L \int\int d\omega_1 d\omega' A_p(\omega_1) A_p(\omega'+\omega-\omega_1) (\hb^{r(\ell)}_{a,s})^\dag(\omega')+i \int^L_0 dz \int d\omega' \hm^{r(\ell)}(z,\omega') A_p(\omega-\omega'),
$
where $\hb^{r(\ell)}_{s,a}$ ($\ha^{r(\ell)}_{s,a}$) are the input (output) annihilation operators for the Stokes and anti-Stokes photons, respectively, in the right (left) fiber spool. $A_p(\omega)$ is the spectral amplitude of the pump in each fiber spool, with $2\pi\int d\omega |A_p(\omega)|^2$ giving the pump-pulse energy; $\alpha(\omega-\omega')$ is determined self-consistently to preserve the commutation relations of the output operators; $\gamma$ is the fiber SFWM coefficient, which we have assumed to be constant; $L$ is the effective length of the fiber spool; and $\hm^{r(\ell)}(z,\omega)$ is the phonon-noise operator accounting for the Raman scattering, which satisfies $[\hm^{r(\ell)}(z,\omega),\hm^{r(\ell)\dag}(z',\omega')]=2\pi
g(\omega) \delta(z-z')\delta(\omega-\omega')$,
where $g(\omega)>0$ is the Raman gain coefficient
\cite{KarDouHau94,RamanMeasured05}. For a phonon bath in
equilibrium at temperature $T$, we have the expectation
$\langle \hm^{r(\ell)\dag}(z,\omega)\hm(z',\omega') \rangle=2\pi
g(\omega) \delta(z-z')\delta(\omega-\omega') n_T(\omega)$, where
$n_T(\omega)=\frac{1}{e^{\hbar |\omega|/k_B T}-1}+\theta(-\omega)$
with $k_B$ the Boltzman constant, and $\theta(\omega)=1$ for
$\omega\ge 0$, and $0$ otherwise.

For the fourfold coincidence measurement depicted in Fig.~\ref{expresults}, the photon-number operators for detectors A, B, C, D are given by $ \hat{n}_M=\sum_{j_M} \eta_M \chi_{j_M}\ha^\dag_{j_M}\ha_{j_M}+\zeta_{M} \hat{d}^\dag_M\hat{d}_M$,
for $M$ = A, B, C, D. Here, $\eta_M$ is the total detection
efficiency taking into account propagation losses and the detector
quantum efficiency. $\eta_{jM}$ is the $j$-th eigenvalue of
the filtering system for detector $M$. $\zeta_M$ measures
the quantum-noise level of the detector as a result of the dark
counts and the after-pulsing counts. $\hat{d}_M$ is a noise operator
obeying $[\hat{d}_M,\hat{d}^\dag_{M'}]=\delta_{M,M'}$. By this
definition, the mean number of dark counts for detector $M$
is then given by the expectation $\zeta_M\langle\hat{d}^\dag_M
\hat{d}_{M'}\rangle$. The bosonic operators
$ \ha_{j_\mathrm{A}(j_\mathrm{B})}=\frac{1}{2\sqrt{2}\pi} \int_{\mathrm{A(B)}} d\omega\left[e^{\frac{i\tau \omega}{2}}\ha^{r}_{s}(\omega)\pm e^{\frac{-i\tau \omega}{2}}  \ha^\ell_s(\omega)\right]\phi_{j_\mathrm{A}(j_\mathrm{B})}(\omega)$, and
    $\ha_{j_\mathrm{C}(j_\mathrm{D})}= \frac{1}{2\pi} \int_{\mathrm{C(D)}} d\omega~\ha^{r(\ell)}_{a} \phi_{j_\mathrm{C}(j_\mathrm{D})}$
where $\tau$ is the amount of signal delay and $``\int_M d\omega''$ represents integral over the detection spectral band of the detector $M$. With $\hat{n}_M$, the positive operator-valued measure
for the detector $M$ to click is calculated to be
$\hat{P}_M=1-:\!\exp(-\hat{n}_M)\!:$,  where ``:~:'' stands for normal ordering of all the embraced
operators. The four-fold coincidence probability is then given by
$\langle :\!\hat{P}_\mathrm{A} \hat{P}_\mathrm{B} \hat{P}_\mathrm{C} \hat{P}_\mathrm{D}\!:\rangle$.

Applying the above theory to the experimental configurations presented
above, we find the predicted visibilities of 17\% (multimode regime) and
72\% (single-mode regime)---in excellent agreement with the experimental
results.  Note that because the theoretical fits shown in Fig.~\ref{expresults} are
generated from the complete theory described above, they require \emph{no
fitting parameter}.  As a result, we conclude that the theories of both
single-mode filtering and SFWM in the presence of noise are able to
accurately model our experiments in both the single-mode and multimode
regimes, and provide an important new tool for the study of
distinguishability in photonic systems.

This research was supported in part by the Defense
Advanced Research Projects Agency (DARPA) under
the Zeno-based Opto-Electronics (ZOE) program (Grant
No. W31P4Q-09-1-0014) and by the United States Air
Force Office of Scientific Research (USAFOSR) (Grant
No. FA9550-09-1-0593).

\bibliographystyle{apsrev4-1}

\begin{thebibliography}{99}

\bibitem{ent}
R. Horodecki, P. Horodecki, M. Horodecki and K. Horodecki, Rev. Mod. Phys. \textbf{81}, 865¨C942 (2009).

\bibitem{decoh1}
W. J. Zurek, Rev. Mod. Phys. \textbf{75}, 715¨C775 (2003).

\bibitem{decoh2}
M. Schlosshauer, Rev. Mod. Phys. \textbf{76}, 1267¨C1305 (2005).

\bibitem{indis}
L. Mandel, Optics Letters \textbf{16}, 23 (1982).

\bibitem{bb84}
C. H. Bennett and G. Brassard, in proc. of the IEEE International Conference on Computers, Systems and Signal Processing, Bangalore, India (1984).

\bibitem{KniLafMil01}
E. Knill, R. Laflamme and G. J. Milburn, Nature \textbf{409}, 46 (2001).

\bibitem{quantummetrology}
V. Giovannetti, S. Lloyd and L. Maccone, Nature Photonics \textbf{5}, 222 (2011).

\bibitem{HuaAltKum10}
Yu-Ping Huang, Joseph B. Altepeter, and Prem Kumar, Phys. Rev. A \textbf{82}, 043826 (2010).

\bibitem{HuaAltKum11}
Yu-Ping Huang, Joseph B. Altepeter, and Prem Kumar, Phys. Rev. A \textbf{84}, 033844 (2011).

\bibitem{ZeiHorWei97}
A. Zeilinger, M. A. Horne, H. Weinfurter and M. Zukowski, Phys. Rev. Lett. \textbf{78}, 3031 (1997).

\bibitem{FioVosSha02}
M. Fiorentino, P. L. Voss, J. E. Sharping and P. Kumar, Photon. Technol. Lett. \textbf{27}, 491 (2002).

\bibitem{Heralded-Single-Photon-SPDC86}
C. K. Hong and L. Mandel, Phys. Rev. Lett. \textbf{56}, 58 (1986).

\bibitem{Single-Photon-PCF05}
J. Fulconis, O. Alibart, W. Wadsworth, P. Russell and J. Rarity, Opt. Express \textbf{13}, 7572 (2005).

\bibitem{Single-Photon-Fiber09}
O. Cohen, J. S. Lundeen, B. J. Smith, G. Puentes, P. J. Mosley, and I. A. Walmsley, Phys. Rev. Lett. \textbf{102}, 123603 (2009).

\bibitem{Single-Photon-PDC-99}
A. V. Sergienko, M. Atat\"ure, Z. Walton, G. Jaeger, B. E. A. Saleh and M. C. Teich, Phys. Rev. A \textbf{60}, R2622 (1999).

\bibitem{PrSp61}
D. Splepian and H. O. Pollak, Bell Syst. Tech. J. \textbf{40}, 43 (1961).

\bibitem{ZhuCav90}
C. Zhu and C. M. Caves, Phys. Rev. A \textbf{42}, 6794 (1990).





\bibitem{SasSuz06}
M. Sasaki and S. Suzuki, Phys. Rev. A \textbf{73}, 043807 (2006).

\bibitem{gisin03}
H. D. Riedmatten, I. Marcikic, W. Tittel, H. Zbinden and N. Gisin, Phys. Rev. A \textbf{67}, 022301 (2003).

\bibitem{zeilingerentswp09}
R. Kaltenbaek, R. Prevedel, M. Aspelmeyer, M. and A. Zeilinger, Phys. Rev. A \textbf{79} 040302, (2009).

\bibitem{rarity07}
J. Fulconis, O. Alibart, J. L. O¡¯Brien, W. J. Wadsworth, and J. G. Rarity, Phys. Rev. Lett. \textbf{99}, 120501 (2007).

\bibitem{takesue07}
H. Takesue, Appl. Phy. Lett. \textbf{90}, 204101 (2007).

\bibitem{HuaAltKum09}
M. A. Hall, J. B. Altepeter and P. Kumar, Opt. Express \textbf{17}, 14558 (2009).

\bibitem{FWM-Raman07}
Q. Lin, F. Yaman and G. P. Agrawal, Phys. Rev. A \textbf{75}, 023803 (2007).

\bibitem{KarDouHau94}
F. X. K\"{a}rtner, D. J. Dougherty, H. A. Haus and E. P. Ippen, J. Opt. Soc. Am. B \textbf{11}, 1267 (1994).

\bibitem{RamanMeasured05}
X. Li, P. Voss, J. Chen, K. F. Lee and P. Kumar, Opt. Express \textbf{13}, 2236 (2005).


\end{thebibliography}

\end{document}